\documentclass[12pt]{article}

\usepackage{mathptmx}       
\usepackage{helvet}         
\usepackage{courier}        
\usepackage{type1cm}        
\usepackage{makeidx}         
\usepackage{graphicx}        
\usepackage{multicol}        
\usepackage[bottom]{footmisc}

\usepackage{amsmath,amssymb}
\usepackage{bm}

\makeindex             

\def\siopK{K}

\def\sio2{\frac{1}{2}}
\def\sio4{\frac{1}{4}}

\def\siobea{\begin{eqnarray}}
\def\sioeea{\end{eqnarray}}
\def\siobes{\begin{eqnarray}}
\def\sioees{\end{eqnarray}}
\def\siobe{\begin{equation}}

\def\sioee{\end{equation}}

\def\sioba{\begin{array}}
\def\sioea{\end{array}}
\def\siobi{\begin{itemize}}
\def\sioei{\end{itemize}}

\def\siodslash{\not{\hbox{\kern-2pt $\partial$}}}

\def\sioeslash{\not{\hbox{\kern-2pt $\epsilon$}}}

\def\sioDslash{\not{\hbox{\kern-4pt $D$}}}

\def\sioAslash{\not{\hbox{\kern-4pt $A$}}}

\def\sioQslash{\not{\hbox{\kern-4pt $Q$}}}

\def\sioWslash{\not{\hbox{\kern-4pt $W$}}}

\def\siopslash{\not{\hbox{\kern-2.3pt $p$}}}

\def\siokslash{\not{\hbox{\kern-2.3pt $k$}}}

\def\sioqslash{\not{\hbox{\kern-2.3pt $q$}}}

\begin{document}

\title{Perturbations of anti-de Sitter black holes\footnote{Research supported in part by the US Department of Energy under grant DE-FG05-91ER40627.}}
\author{George Siopsis}
\author{George Siopsis\footnote{E-mail: siopsis@tennessee.edu}\\
\em Department of Physics
and Astronomy, \\
\em The University of Tennessee, Knoxville, \\
\em TN 37996 - 1200, USA.
}
\date{}
%
%
\maketitle

\vspace{-3.5in}\hfill UTHET-10-0101\vspace{3.5in}


\abstract{
I review perturbations of black holes in asymptotically anti-de Sitter space. I show how the quasi-normal modes governing these perturbations can be calculated analytically and discuss the implications on the hydrodynamics of gauge theory fluids per the AdS/CFT correspondence. I also discuss phase transitions of hairy black holes with hyperbolic horizons and the dual superconductors emphasizing the analytical calculation of their properties.
}

\vspace{2in}

\noindent{\sl Prepared for the proceedings of the 5th Aegean Summer School}
From Gravity to Thermal Gauge Theories: the AdS/CFT Correspondence, {\sl Milos,
Greece, September 2009.}

\newpage

\section{Introduction}
\label{siosec:1}
The perturbations of a black hole are governed by quasi-normal modes (QNMs).
The latter are typically obtained by solving
a wave equation for small fluctuations in the black hole background subject to the conditions that the
flux be
ingoing at the horizon and
outgoing at asymptotic infinity.
These boundary conditions in general lead to a
discrete spectrum of complex frequencies whose
imaginary part determines the decay time of the small fluctuations
\siobe \Im \omega = \frac{1}{\tau}\sioee
There is a vast literature on quasi-normal modes and I make no attempt to review it here. Instead, I concentrate on obtaining analytic expressions for QNMs of black hole perturbations in asymptotically AdS space. One can rarely obtain analytic expressions in closed form. Instead, I discuss techniques which allow one to calculate the spectrum perturbatively in an asymptotic regime (high or low overtones). For high overtones, the frequencies at leading order are proportional to the radius of the horizon. For low overtones, one in general obtains an additional frequency which is inversely proportional to the horizon radius. Thus for large black holes there is a gap between the lowest frequency and the rest of the spectrum of QNMs.
I pay special attention to the lowest frequencies because they govern the behavior of the gauge theory fluid on the boundary according to the AdS/CFT correspondence.
The latter may have experimental consequencies pertaining to the formation of the quark-gluon plasma in heavy ion collisions. Moreover, I discuss phase transitions to hairy black holes which correspond to a dual superconducting phase. I concentrate on the case of black holes with a hyperbolic horizon because their properties can be understood analytically.

In section \ref{siosec:2} I discuss scalar, gravitational and electromagnetic perturbations of an AdS Schwarzschild black hole analytically calculating the QNM spectrum in the high frequency regime. In section \ref{siosec:3} I calculate the QNM spectrum analytically in the low frequency regime and discuss its relevance to the hyrdodynamic behavior of the dual gauge theory fluid on the boundary.
In section \ref{siosec:4} I introduce hairy black holes and discuss their phase transition in the case of a hyperbolic horizon which can be understood analytically. The dual gauge theory corresponds to a superconductor whose properties can be calculated via electromagnetic perturbations. Finally, I conclude with section \ref{siosec:5}.



\section{Perturbations}
\label{siosec:2}

In this section I discuss scalar, gravitational and electromagnetic perturbations of an AdS Schwarzschild black hole in $d$ dimensions analytically calculating the QNM spectrum in the high frequency regime. Low overtones will be discussed in the next section.

The metric of an AdS Schwarzschild black hole is
\siobe\label{siometric}
ds^2=-\left(\frac{r^2}{l^2}+\siopK-\frac{2\mu}{r^{d-3}}\right)dt^2+\frac{dr^2}{\frac{r^2}{l^2}+\siopK-\frac{2\mu}{r^{d-3}}}+r^2d\Sigma^2_{\siopK,d-2}
\sioee
I shall choose units so that the AdS radius $l=1$.
The
horizon radius and Hawking temperature are, respectively,
\siobe
2\mu=r_+^{d-1}\left( 1+\frac{\siopK}{r_+^2}\right)~,~~~~T_H=\frac{(d-1)r_+^2+\siopK(d-3)}{4\pi r_+}
\sioee
The mass and entropy of the hole are, respectively,
\siobe\label{sioBH}
M=(d-2)(\siopK+r_+^2)\frac{r_+^{d-3}}{16\pi G} Vol(\Sigma_{\siopK,d-2})~,~~~ S=\frac{r_+^{d-2}}{4G} Vol(\Sigma_{\siopK,d-2})
\sioee
The parameter $K$ determines the curvature of the horizon and the boundary of AdS space. For $K=0,+1,-1$ we have, respectively, a flat ($\mathbb{R}^{d-2}$), spherical ($\mathbb{S}^{d-2}$) and hyperbolic ($\mathbb{H}^{d-2} /\Gamma$, topological black hole, where $\Gamma$ is a discrete group of isometries) horizon (boundary).

The harmonics on $\Sigma_{\siopK,d-2}$ satisfy
\siobe
\left(\nabla^2 + k^2\right){\mathbb T}=0
\sioee
For $\siopK = 0$, $k$ is the momentum; for
$\siopK = +1$, the eigenvalues are quantized,
\siobe
k^2=l(l+d-3)-\delta
\sioee
whereas for $\siopK = -1$,
%
\siobe
k^2=\xi^2+\left(\frac{d-3}{2}\right)^2+\delta
\sioee
where $\xi$ is discrete for non-trivial $\Gamma$.
$\delta =0,1,2$ for scalar, vector, or tensor perturbations, respectively.

According to the AdS/CFT correspondence,
QNMs of AdS black holes are expected to
correspond to perturbations of the dual Conformal Field Theory (CFT) on the boundary.
The
establishment of
such a correspondence is hindered by difficulties in solving the wave equation
governing the various types of perturbation.
In three dimensions one obtains a hypergeometric equation which leads to
explicit analytic expressions for the QNMs \cite{siob-CL,siob-BSS}.
In five dimensions one obtains a Heun equation and a derivation of analytic expressions for QNMs is no longer possible.
On the other hand, numerical results exist in four, five and seven dimensions
\cite{siob-HH,siob-Star,siob-Kono}.

\subsection{Scalar perturbations}


To find the asymptotic form of QNMs, we need to find an approximation to the wave equation valid in the
high frequency regime.
In three dimensions the resulting wave equation will be an exact equation (hypergeometric equation).
In five dimensions, I shall turn the Heun equation into a hypergeometric equation
which will lead to
an analytic expression for the asymptotic form of QNM frequencies
in agreement
with numerical results.

\subsubsection{AdS$_3$}

In three dimensions the wave equation for a massless scalar field is
\siobe
\frac{1}{r}\partial_r \left( r^3 \left( 1- \frac{r_+^2}{r^2}\right) \partial_r \Phi\right) -\frac{1}{r^2 - r_+^2 }\partial_t^2 \Phi + \frac{1}{r^2}\partial_x^2 \Phi =0
\sioee
Writing the wavefunction in the form
\siobe
\Phi = e^{i(\omega t-px) }\Psi (y) ,\ \ \ \ \ y = \frac{r_+^2}{r^2}
\sioee
the wave function becomes
\siobe
y^2 (y-1)\left( (y-1) \Psi' \right)'
+ \hat\omega^2\, y\Psi +\hat p^2\, y(y-1)\Psi
=0
\sioee
to be solved in the interval $0<y<1$, where
\siobe
\hat\omega = \frac{\omega}{2r_+} = \frac{\omega}{4\pi T_H},\ \ \ \hat p = \frac{p}{2r_+} = \frac{p}{4\pi T_H}\;.
\sioee
For QNMs, we are interested in the
solution
\siobe \Psi (y) = y(1-y)^{i\hat\omega} {}_2F_1 (1+i(\hat\omega + \hat p), 1+i(\hat\omega - \hat p); 2; y)\sioee
which vanishes at the boundary ($y\to 0$).
Near the horizon ($y\to 1$), we obtain a mixture of ingoing and outgoing waves,
\siobe \Psi \sim A_+ (1-y)^{-i\hat\omega} + A_-(1-y)^{+i\hat\omega}\ \ , \ \ \ \
A_\pm =
\frac{\Gamma(\pm 2i\hat\omega)}{\Gamma(1\pm i(\hat\omega + \hat p))\Gamma(1\pm i(\hat\omega - \hat p))}\nonumber\sioee
Setting $A_- =0$, we deduce the quasi-normal frequencies
\siobe \hat\omega = \pm \hat p  -in\quad,\quad n=1,2,\dots \sioee
which form a discrete spectrum of complex frequencies with $\Im\hat\omega < 0$.



\subsubsection{AdS$_5$}

Restricting attention to the case of a large black hole, the massless scalar wave equation reads
\siobe
\frac{1}{r^3}\partial_r (r^5\, f(r)\, \partial_r \Phi) -\frac{1}{ r^2\, f(r) }\partial_{t}^2\Phi - \frac{1}{r^2}\; \vec\nabla^2\Phi = 0
\ \ , \ \ \ \ \ f(r) = 1- \frac{r_+^4}{r^4}
\sioee
Writing the solution in the form
\siobe 
\Phi = e^{i(\omega t - \vec p\cdot \vec x)} \Psi (y)
\ \ , \ \ \ \
y = \frac{r^2}{r_+^2} 
\sioee
the radial wave equation becomes
\siobe
(y^2-1)\left( y(y^2-1) \Psi' \right)' + \left(\frac{\hat\omega^2}{4}\, y^2 - \frac{\hat p^2}{4}\, (y^2-1)\right)\Psi = 0
\sioee
%
For QNMs, we are interested
in the analytic solution which vanishes at the boundary and behaves as an ingoing wave at the horizon.
The wave equation contains an additional
(unphysical) singularity at $y=-1$, at which the wavefunction behaves as
$\Psi \sim (y+1)^{\pm \hat\omega /4}$.
Isolating the behavior of the wavefunction near the singularities $y=\pm 1$,
\siobe
\Psi (y) = (y-1)^{-i\hat\omega/4} (y+1)^{\pm\hat\omega/4} F_\pm (y)
\sioee
we shall obtain two sets of modes with the same $\Im\hat\omega$, but opposite
$\Re\hat\omega$.

$F_\pm (y)$ satisfies the Heun equation
\siobea
y(y^2-1) F\pm'' + \left\{ \left( 3- \frac{i\pm 1}{2}\, \hat\omega \right) y^2 - \frac{i \pm 1}{2}\, \hat\omega y -1 \right\} F_\pm' & & \nonumber\\
+ \left\{ \frac{\hat\omega}{2}\left( \pm \frac{i\hat\omega}{4} \mp 1-i\right) y - (i\mp 1)\frac{\hat\omega}{4} - \frac{\hat p^2}{4} \right\}\; F_\pm &=& 0 
\sioeea
to be solved in a region in the complex $y$-plane
containing $|y|\ge 1$ which
includes the physical regime $r> r_+$.

For large $\hat\omega$, the constant terms in the polynomial coefficients of $F'$ and $F$ are small compared with the other terms,
therefore they may be dropped.
The wave equation may then be approximated by a hypergeometric equation
\siobe
(y^2-1) F_\pm'' + \left\{ \left( 3- \frac{i\pm 1}{2}\, \hat\omega \right) y - \frac{i \pm 1}{2}\, \hat\omega \right\} F_\pm'
+ \frac{\hat\omega}{2}\left( \pm \frac{i\hat\omega}{4} \mp 1-i\right)\; F_\pm =0
\sioee
in the asymptotic limit of large frequencies $\hat\omega$.
The acceptable solution is
\siobe F_0(x) = {}_2F_1 ( a_+, a_-; c; (y+1)/2)
\ \ , \ \ \ a_\pm = 1-{\textstyle{\frac{i \pm 1}{4}}}\,\hat\omega\pm 1
\quad,\quad c = {\textstyle{\frac{3}{2}}}\pm {\textstyle{\frac{1}{2}}}\,\hat\omega\sioee
For proper behavior at the boundary ($y\to\infty$), we demand that $F$ be a {\em polynomial},
which leads to the condition
\siobe a_+ = -n \ \ , \ \ n = 1,2,\dots\sioee
Indeed, it implies that $F$ is a polynomial of order $n$, so as $y\to\infty$,
$F\sim y^n
\sim y^{-a_+}$
and $\Psi \sim y^{-i\hat\omega/4} y^{\pm\hat\omega/4} y^{-a_+} \sim y^{-2}$,
as expected.

We deduce the quasi-normal frequencies \cite{siob-MS4}
\siobe \hat\omega = \frac{\omega}{4\pi T_H} = 2n(\pm 1-i) \sioee
in agreement with numerical results.


It is perhaps worth mentioning that these frequencies may also be deduced by a simple monodromy argument \cite{siob-MS4}.
Considering the monodromies around the singularities,
if the wavefunction has no singularities other than $y=\pm 1$,
the contour around $y=+1$ may be unobstructedly deformed into the contour
around $y=-1$, which yields
\siobe \mathcal{M} (1) \mathcal{M} (-1) = 1\sioee
Since the respective monodromies are
$\mathcal{M} (1) = e^{\pi \hat\omega /2}$ and $\mathcal{M} (-1) = e^{\mp i\pi \hat\omega /2}$,
using $\Im\hat\omega < 0$, we deduce
$\hat\omega = 2n(\pm 1-i)$, in agreement with our result above.

\subsection{Gravitational perturbations}

Next I consider gravitational perturbations. For definiteness, I concentrate on the case of spherical black holes ($\siopK =+1$).
%
I shall derive analytic expressions for QNMs \cite{siob-NS} including
first-order corrections \cite{siob-MNS}.
The results 
are in good agreement with results of
numerical analysis \cite{siob-CKL}.
Extension to other forms of the horizon is straightforward \cite{siob-AS1}.

The radial wave equation
for gravitational perturbations in the black-hole
background~(\ref{siometric})
can be cast into a Schr\"odinger-like form,
\siobe\label{siosch}
  -\frac{d^2\Psi}{dr_*^2}+V[r(r_*)]\Psi =\omega^2\Psi \;,
\sioee
in terms of the tortoise coordinate defined by
\siobe\label{siotortoise}
  \frac{dr_*}{dr} = \frac{1}{f(r)}\;.
\sioee
The potential $V$ for the various types of perturbation has been found by Ishibashi and Kodama \cite{siob-IK}.
For tensor, vector and scalar perturbations, one obtains, respectively,
\siobe\label{sioeqVT} V_{\mathsf{T}} (r) = f(r) \left\{ \frac{\ell (\ell +d-3)}{r^2} + \frac{(d-2)(d-4) f(r)}{4r^2} + \frac{(d-2) f'(r)}{2r} \right\} \sioee
\siobe\label{sioeqVV} V_{\mathsf{V}}(r) = f(r) \left\{ \frac{\ell (\ell +d-3)}{r^2} + \frac{(d-2)(d-4) f(r)}{4r^2} - \frac{r f'''(r)}{2(d-3)} \right\} \sioee
\siobea\label{sioeqVS} V_{\mathsf{S}}(r) &=& \frac{f(r)}{4r^2} \left[ \ell (\ell +d-3) - (d-2) + \frac{(d-1)(d-2)\mu}{r^{d-3}} \right]^{-2} \nonumber\\
&\times& \Bigg\{ \frac{d(d-1)^2(d-2)^3 \mu^2}{R^2r^{2d-8}}
- \frac{6(d-1)(d-2)^2(d-4)[\ell (\ell+d-3) - (d-2)] \mu}{R^2r^{d-5}}\nonumber\\
&& + \frac{(d-4)(d-6)[\ell (\ell+d-3) - (d-2)]^2 r^2}{R^2} +
\frac{2(d-1)^2(d-2)^4 \mu^3}{r^{3d-9}}\nonumber\\
&& + \frac{4(d-1)(d-2)(2d^2-11d+18)[\ell (\ell+d-3) - (d-2)]\mu^2}{r^{2d-6}}\nonumber\\
&& + \frac{(d-1)^2(d-2)^2(d-4)(d-6)\mu^2}{r^{2d-6}}
- \frac{6(d-2)(d-6)[\ell (\ell+d-3) - (d-2)]^2 \mu}{r^{d-3}}\nonumber\\
&& - \frac{6(d-1)(d-2)^2(d-4)[\ell (\ell+d-3) - (d-2)] \mu}{r^{d-3}}\nonumber\\
&& + 4 [\ell (\ell+d-3) - (d-2)]^3 + d(d-2) [\ell (\ell+d-3) - (d-2)]^2 \Bigg\} \nonumber\sioeea
Near the black hole singularity ($r\sim 0$),
\siobe\label{sioeqVT0} V_{\mathsf{T}} = -\frac{1}{4r_*^2}+\frac{\mathcal{A}_{\mathsf{T}} }{[-2(d-2)\mu]^{\frac{1}{d-2}}} r_*^{-\frac{d-1}{d-2}} + \dots \, , \ \ \ \
  \mathcal{A}_{\mathsf{T}} = \frac{(d-3)^2}{2(2d-5)}+\frac{\ell(\ell+d-3)}{d-2},
\sioee
\siobe\label{sioeqVV0} V_{\mathsf{V}} = \frac{3}{4r_*^2}+\frac{\mathcal{A}_{\mathsf{V}} }{[-2(d-2)\mu]^{\frac{1}{d-2}}} r_*^{-\frac{d-1}{d-2}} + \dots \ \ , \ \ \ \
\mathcal{A}_{\mathsf{V}} = \frac{d^2-8d+13}{2(2d-15)} + \frac{\ell (\ell +d-3)}{d-2}\sioee
and
\siobe\label{sioeqVS0}
  V_{\mathsf{S}} =  -\frac{1}{4r_*^2}+\frac{\mathcal{A}_{\mathsf{S}} }{[-2(d-2)\mu]^{\frac{1}{d-2}}} r_*^{-\frac{d-1}{d-2}} + \dots \, ,
\sioee
where
\siobe\label{sioeqVS0a}
  \mathcal{A}_{\mathsf{S}} = \frac{ (2d^3-24d^2+94d-116)}{4(2d-5)(d-2)}+\frac{ (d^2-7d+14)[ \ell(\ell+d-3)-(d-2)]}{(d-1)(d-2)^2}
\sioee
I have included only the terms which contribute to the order I am interested in.
The behavior of the potential near the origin may be summarized by
\siobe\label{sioeqV0} V= \frac{j^2 -1}{4r_*^2}+\mathcal{A}\, r_*^{-\frac{d-1}{d-2}} + \dots \sioee
where $j=0$ ($2$) for scalar and tensor (vector) perturbations.

On the other hand, near the boundary (large $r$),
\siobe V = \frac{j_\infty^2-1}{4(r_*-\bar r_*)^2} + \dots \ \ , \ \ \ \ \bar r_* = \int_0^\infty \frac{dr}{f(r)} \sioee
where $j_\infty = d-1$, $d-3$ and $d-5$ for tensor, vector and scalar perturbations,
respectively.

After rescaling the tortoise coordinate $(z=\omega r_*)$,
the wave equation to first order becomes
\siobe\label{siowe-h}
  \left( \mathcal{H}_0+\omega^{-\frac{d-3}{d-2}} \, \mathcal{H}_1 \right) \Psi =0,
\sioee
where
\siobe\label{sioH0-H1}
  \mathcal{H}_0= \frac{d^2}{dz^2}-\left[\frac{j^2-1}{4z^2}-1\right]\ \ ,\ \ \mathcal{H}_1=-\mathcal{A}
\; z^{-\frac{d-1}{d-2}}.
\sioee
By treating $\mathcal{H}_1$ as a perturbation, one may expand the wave function
\siobe\label{sioexpandwf}
  \Psi(z)=\Psi_0(z)+\omega^{-\frac{d-3}{d-2}} \, \Psi_1(z)+\dots
\sioee
and solve the wave equation perturbatively.

The zeroth-order wave equation,
\siobe\label{siowe-0}
  \mathcal{H}_0\Psi_0(z)=0,
\sioee
may be solved in terms of Bessel functions,
\siobe\label{siosoln_0}
  \Psi_0(z)=A_1\sqrt{z}\, J_{\frac{j}{2}}(z)+A_2 \sqrt{z}\, N_{\frac{j}{2}}(z).
\sioee
For large $z$, it behaves as
\siobea \label{siosoln-0-0}
  \Psi_0(z)&\sim&  \sqrt{\frac{2}{\pi}}\left[A_1\cos(z-\alpha_+)+A_2\sin(z-\alpha_+)\right]\nonumber\\
  &=&\frac{1}{\sqrt{2\pi}}(A_1-iA_2)e^{-i\alpha_+}e^{iz} + \frac{1}{\sqrt{2\pi}}(A_1+iA_2)e^{+i\alpha_+}e^{-iz}\nonumber
\sioeea
where $\alpha_\pm = \frac{\pi}{4}(1\pm j)$.

At the boundary ($r\to\infty$),
the wavefunction ought to vanish,
therefore the acceptable solution is
\siobe\label{siosoln-0-infty}
  \Psi_0(r_*) = B\sqrt{\omega(r_*-\bar r_*)}\; J_{\frac{j_\infty}{2}}(\omega (r_*-\bar r_*))
\sioee
Indeed, $\Psi \to 0$ as $r_*\to \bar r_*$, as desired.

Asymptotically (large $z$), it behaves as
\siobe\label{sioeq54} \Psi(r_*) \sim \sqrt{\frac{2}{\pi}}\, B\cos\left[ \omega(r_*-\bar r_*)+\beta\right] \ , \ \ \ \
\beta =\frac{\pi}{4}(1+ j_\infty) \sioee
This ought to be matched to the asymptotic form of the wavefunction in the vicinity of the black-hole singularity along the Stokes line $\Im z = \Im (\omega r_*) = 0$.
This leads to a
constraint on the coefficients $A_1,\ A_2$,
\siobe \label{sioconstraint_1}
  A_1\tan(\omega \bar r_* -\beta -\alpha_+)-A_2=0.
\sioee
By imposing the boundary condition at the horizon
\siobe  \Psi(z) \sim e^{iz}\ \ , \ \ \ \ z\to -\infty\label{siobc-0}\ ,
\sioee
one obtains a second constraint.
To find it,
one needs to analytically continue the wavefunction near the
black hole singularity ($z=0$) to negative values of $z$.
A rotation of $z$ by $-\pi$ corresponds to a rotation by $-\frac{\pi}{d-2}$ near the origin in the complex $r$-plane.
Using the known behavior of Bessel functions
\siobe\label{sioeqBrot} J_\nu (e^{-i\pi} z) = e^{-i\pi\nu} J_\nu (z) \ , \ \ \ \
N_\nu (e^{-i\pi} z) = e^{i\pi\nu} N_\nu (z) - 2i\cos \pi\nu\, J_\nu (z)\sioee
for $z<0$ the wavefunction changes to
\siobe\label{siosoln_0r}
  \Psi_0(z)= e^{-i\pi(j+1)/2} \sqrt{-z}\, \left\{ \left[ A_1 -i (1+e^{i\pi j}) A_2 \right]\, J_{\frac{j}{2}}(-z)+A_2 e^{i\pi j} \, N_{\frac{j}{2}}(-z) \right\}
\sioee
whose asymptotic behavior is given by
\siobe \Psi \sim \frac{e^{-i\pi(j+1)/2}}{\sqrt{2\pi}} \left[ A_1-i(1+2e^{j\pi i}) A_2\right]\, e^{-iz}+\frac{e^{-i\pi(j+1)/2}}{\sqrt{2\pi}} \left[ A_1-iA_2\right]\, e^{iz} \sioee
Therefore one obtains a second constraint
\siobe\label{sioconstraint_2}
  A_1 -i(1+2e^{j\pi i}) A_2 = 0\ \ .
\sioee
The two constraints are compatible provided
\siobe\label{sioeqcomp}
  \left| \begin{array}{cc}  1 &  -i(1+2e^{j\pi i}) \\
                          \tan(\omega \bar r_*-\beta-\alpha_+) & -1 \end{array}   \right| = 0
\sioee
which yields the quasi-normal frequencies \cite{siob-NS}
\siobe
  \omega \bar r_* =\frac{\pi}{4} (2+j+ j_\infty)-\tan^{-1} \frac{i}{1+2e^{j\pi i}} +n\pi
\sioee

The first-order correction to the above asymptotic expression may be found by standard perturbation theory \cite{siob-MNS}.
%
%
%
To first order, the wave equation becomes 
\siobe\label{sio1stwe1}
  \mathcal{H}_0\Psi_1+\mathcal{H}_1\Psi_0=0
\sioee
The solution is
\siobe\label{siosoln_1}
  \Psi_1(z) = \sqrt{z}\, N_{\frac{j}{2}}(z)\int_0^z dz'\frac{\sqrt{z'}\, J_{\frac{j}{2}}(z')
\mathcal{H}_1\Psi_0(z') }{ \mathcal{W} } -  \sqrt{z}\, J_{\frac{j}{2}}(z)\int_0^z dz'\frac{\sqrt{z'}\, N_{\frac{j}{2}}(z') \mathcal{H}_1\Psi_0(z') }{ \mathcal{W} }
\sioee
where $\mathcal{W} = 2/\pi$ is the Wronskian.

The wavefunction to first order reads
\siobe\label{siosoln1st0}
  \Psi(z)=\left\{A_1[1-b(z)] -A_2a_2(z)\right\}\sqrt{z} J_{\frac{j}{2}}(z) +\left\{A_2[1+b(z)]+A_1a_1(z)\right\}\sqrt{z} N_{\frac{j}{2}}(z)
\sioee
where
\siobea
  a_1(z) &=& \frac{\pi\mathcal{A}}{2} \, \omega^{-\frac{d-3}{d-2}}\, \int_0^z dz'\;{z'}^{-\frac{1}{d-2}}J_{\frac{j}{2}}(z') J_{\frac{j}{2}}(z') \nonumber\\
  a_2(z) &=& \frac{\pi\mathcal{A}}{2} \, \omega^{-\frac{d-3}{d-2}}\,  \int_0^z dz'\;{z'}^{-\frac{1}{d-2}}N_{\frac{j}{2}}(z') N_{\frac{j}{2}}(z') \nonumber\\
  b(z) &=& \frac{\pi\mathcal{A}}{2} \, \omega^{-\frac{d-3}{d-2}}\,  \int_0^z dz'\;{z'}^{-\frac{1}{d-2}}J_{\frac{j}{2}}(z') N_{\frac{j}{2}}(z') \nonumber
\sioeea
and $\mathcal{A}$ depends on the type of perturbation.

Asymptotically, it behaves as
\siobe\label{sio1stsoln}
  \Psi(z)\sim \sqrt{\frac{2}{\pi}}\, [A_1' \cos(z-\alpha_+)+ A_2' \sin(z-\alpha_+)]\ ,
\sioee
where
\siobe\label{sioeq91} A_1' = [1-\bar b]A_1-\bar a_2 A_2\ \ , \ \ \ \
A_2' = [1+\bar b]A_2+\bar a_1 A_1\sioee
and I introduced the notation
\siobe \bar a_1 = a_1(\infty)\ \ , \ \ \ \ \bar a_2 = a_2(\infty)\ \ , \ \ \ \ \bar b = b(\infty) \ . \sioee
The first constraint is modified to
\siobe\label{sionewconstr1} A_1' \tan (\omega \bar r_* -\beta -\alpha_+) - A_2' = 0\sioee
Explicitly,
\siobe\label{sionewconstr1a} [ (1-\bar b)\tan (\omega \bar r_* -\beta -\alpha_+)- \bar a_1 ]A_1 -[1+\bar b +\bar a_2 \tan (\omega \bar r_* -\beta -\alpha_+)]A_2 = 0\sioee
To find the second constraint to first order,
one needs to approach the horizon.
This entails a rotation by $-\pi$ in the $z$-plane.
Using
\siobea a_1 (e^{-i\pi} z) &=& e^{-i\pi \frac{d-3}{d-2}} e^{-i\pi j} a_1 (z)\ , \nonumber\\
a_2 (e^{-i\pi} z) &=& e^{-i\pi \frac{d-3}{d-2}} \left[ e^{i\pi j} a_2(z)
- 4 \cos^2 \frac{\pi j}{2} a_1(z) - 2i (1+e^{i\pi j} ) b (z) \right]\ , \nonumber\\
b (e^{-i\pi} z) &=& e^{-i\pi \frac{d-3}{d-2}} \left[ b(z) -i (1+e^{-i\pi j}) a_1(z) \right]\nonumber\sioeea
in the limit $z\to -\infty$ one obtains
\siobe\label{siosoln1st0r}
\Psi(z) \sim -i e^{-ij\pi/2} B_1 \cos(-z-\alpha_+)
-i e^{ij\pi/2} B_2\sin(-z-\alpha_+)
\sioee
where
\siobea
B_1 &=& 
   A_1 -A_1e^{-i\pi\frac{d-3}{d-2}}[{\bar b}-i(1+e^{-i\pi j}){\bar a}_1]
\nonumber \\
& & -A_2e^{-i\pi\frac{d-3}{d-2}} \left[ e^{+i\pi j}{\bar a}_2-4\cos^2\frac{\pi j}{2}{\bar a}_1-2i(1+e^{+i\pi j}){\bar b} \right] \nonumber\\
  & & -i (1+e^{i\pi j})\left[ A_2 +A_2e^{-i\pi\frac{d-3}{d-2}}[{\bar b}-i(1+e^{-i\pi j}){\bar a}_1]+A_1 e^{-i\pi\frac{d-3}{d-2}} e^{-i\pi j}{\bar a}_1 \right] \nonumber\\
B_2 &=& A_2+A_2e^{-i\pi\frac{d-3}{d-2}}[{\bar b}-i(1+e^{-i\pi j}){\bar a}_1]+A_1e^{-i\pi\frac{d-3}{d-2}}e^{-i\pi j}{\bar a}_1\nonumber
\sioeea
Therefore the second constraint to first order reads
\siobe\label{sionewconstr2} [1-e^{-i\pi\frac{d-3}{d-2}}(i
\bar a_1 +\bar b)]A_1 -[i(1+2e^{i\pi j})+e^{-i\pi\frac{d-3}{d-2}}((1+e^{i\pi j})\bar a_1 +e^{i\pi j} \bar a_2-i\bar b)
]A_2 = 0 \sioee
Compatibility of the two first-order constraints yields
\siobe\label{sioeq99}
  \left| \begin{array}{cc}  1+\bar b+\bar a_2\tan (\omega \bar r_* -\beta -\alpha_+) & i(1+2e^{i\pi j})+e^{-i\pi\frac{d-3}{d-2}}((1+e^{i\pi j})\bar a_1 +e^{i\pi j} \bar a_2-i\bar b) \\
                          (1-\bar b)\tan (\omega \bar r_* -\beta -\alpha_+)- \bar a_1 & 1-e^{-i\pi\frac{d-3}{d-2}}(i
\bar a_1 +\bar b) \end{array}   \right| = 0
\sioee
leading to the first-order expression for quasi-normal frequencies,
\siobea\label{sioeqo1st}
\omega {\bar r}_* &=& \frac{\pi}{4}(2+j+j_{\infty}) +\frac{1}{2i}\ln 2+n\pi \nonumber\\
   & & -\frac{1}{8}\left\{ 6i\bar b -2i e^{-i\pi\frac{d-3}{d-2}} \bar b  -9\bar a_1+e^{-i\pi\frac{d-3}{d-2}}{\bar a}_1 +{\bar a}_2 - e^{-i\pi\frac{d-3}{d-2}}{\bar a}_2 \right\}
\nonumber\sioeea
where
\siobea\label{sioeqab} \bar a_1 &=& \frac{\pi\mathcal{A}}{4} \left(\frac{n\pi}{2\bar r_*}\right)^{-\frac{d-3}{d-2}} \frac{\Gamma(\frac{1}{d-2})\Gamma(\frac{j}{2}+\frac{d-3}{2(d-2)})}{\Gamma^2(\frac{d-1}{2(d-2)})\Gamma(\frac{j}{2}+\frac{d-1}{2(d-2)})}\nonumber \\
\bar a_2 &=& \left[ 1+2\cot \frac{\pi (d-3)}{2(d-2)} \cot \frac{\pi}{2} \left( -j+\frac{d-3}{d-2}\right) \right]\bar a_1 \nonumber \\
\bar b &=& -\cot \frac{\pi (d-3)}{2(d-2)}\ \bar a_1 \nonumber\sioeea
Thus the first-order correction is $\sim \mathcal{O} (n^{-\frac{d-3}{d-2}})$.

%
The above analytic results are in good agreement with numerical results \cite{siob-CKL}
(see ref.~\cite{siob-MNS} for a detailed comparison).

\subsection{Electromagnetic perturbations}

The
electromagnetic potential in four dimensions is
\siobe\label{sioV-V}
  V_{\mathsf{EM}} =\frac{\ell(\ell+1)}{r^2}f(r).
\sioee
Near the origin,
\siobe\label{sioeqVEMr}
  V_{\mathsf{EM}} =\frac{j^2-1}{4r_*^2}+\frac{\ell(\ell+1)r_*^{-3/2}}{2\sqrt{-4\mu}}+\dots \ ,
\sioee
where $j=1$. Therefore a vanishing potential to zeroth order is obtained.
To calculate the QNM spectrum one needs to include first-order corrections from the outset.
Working as with gravitational perturbations, one obtains the
QNMs
\siobe\label{sioeqEMo1}
\omega {\bar r}_* = n\pi -\frac{i}{4}\ln n+\frac{1}{2i}\ln\left( 2(1+i){\cal A}\sqrt{\bar r_*}\right) \ , \ \ \ \ \mathcal{A}
= \frac{\ell(\ell+1)}{2\sqrt{-4\mu}}
\sioee
Notice that the first-order correction behaves as $\ln n$, a fact which may be associated with gauge invariance.

As with gravitational perturbations, the above analytic results are in good agreement with numerical results \cite{siob-CKL}
(see ref.~\cite{siob-MNS} for a detailed comparison).


\section{Hydrodynamics}
\label{siosec:3}


There is a correspondence between
$\mathcal{N}=4$ Super Yang-Mills (SYM) theory in the large $N$ limit
and type-IIB string theory in $\mathrm{AdS}_5\times \mathrm{S^5}$ (AdS/CFT correspondence).
In the low energy limit, string theory is
reduced to classical supergravity and the AdS/CFT correspondence allows one to calculate all
gauge field-theory correlation functions in the strong coupling limit leading to non-trivial predictions on the behavior of gauge theory fluids.
For example,
the entropy of $\mathcal{N}=4$ SYM
theory in the limit of large
't Hooft coupling is precisely 3/4 its value in the zero
coupling limit.

The long-distance (low-frequency) behavior of any
interacting theory at finite temperature must be described by fluid mechanics 
(hydrodynamics).
This leads to a
universality in physical properties because
hydrodynamics implies very precise constraints on
correlation functions of conserved currents and
the stress-energy tensor.
Their correlators are
fixed once a few transport coefficients are 
known.

\subsection{Vector perturbations}

I start with vector perturbations and work in the $d$-dimensional Schwarzschild background (\ref{siometric}) with $\siopK=+1$ (spherical horizon and boundary). It is convenient to introduce
the coordinate \cite{siob-ego}
\siobe\label{sioeqru} u = \left( \frac{r_+}{r} \right)^{d-3} \sioee
The wave equation becomes
\siobe\label{sioeq13} -(d-3)^2 u^{\frac{d-4}{d-3}}\hat f(u) \left( u^{\frac{d-4}{d-3}}\hat f(u) \Psi' \right)' +\hat V_{\mathsf{V}} (u)\Psi = \hat\omega^2 \Psi  \ \ , \ \ \ \
\hat\omega = \frac{\omega}{r_+}\sioee
where prime denotes differentiation with respect to $u$ and I have defined
\siobe\label{sioeq14} \hat f(u) \equiv \frac{f(r)}{r^2} = 1- u^{\frac{2}{d-3}} \left( u- \frac{1 - u}{r_+^2} \right) \sioee
\siobe\label{sioeq15} \hat V_{\mathsf{V}} (u) \equiv \frac{V_{\mathsf{V}}}{r_+^2} = \hat f(u) \left\{ \hat L^2 + \frac{(d-2)(d-4)}{4} u^{-\frac{2}{d-3}}\hat f(u) - \frac{(d-1)(d-2)\left( 1+ \frac{1}{r_+^2} \right)}{2} u\right\} \sioee
where
$\hat L^2 = \frac{\ell (\ell +d-3)}{r_+^2} $.

First I consider the large black hole limit $r_+ \to\infty$ keeping $\hat\omega$ and $\hat L$ fixed~(small).
Factoring out the behavior at the horizon ($u=1$)
\siobe \Psi = (1-u)^{-i \frac{\hat\omega}{d-1}} F(u) \sioee
the wave equation simplifies to
\siobe\label{siosch2} \mathcal{A} F'' + \mathcal{B}_{\hat\omega} F' + \mathcal{C}_{\hat\omega , \hat L} F = 0 \sioee
where
\siobea \mathcal{A} &=& - (d-3)^2 u^{\frac{2d-8}{d-3}} (1-u^{\frac{d-1}{d-3}}) \nonumber\\
\mathcal{B}_{\hat\omega} &=& - (d-3) [ d-4-(2d-5)u^{\frac{d-1}{d-3}}]u^{\frac{d-5}{d-3}} - 2(d-3)^2 \frac{i\hat\omega}{d-1}\frac{u^{\frac{2d-8}{d-3}} (1-u^{\frac{d-1}{d-3}})}{1-u} \nonumber\\
\mathcal{C}_{\hat\omega , \hat L} &=& \hat L^2 + \frac{(d-2)[d-4-3(d-2)u^{\frac{d-1}{d-3}}]}{4}u^{-\frac{2}{d-3}} \nonumber\\
& & - \frac{\hat\omega^2}{1-u^{\frac{d-1}{d-3}}} + (d-3)^2 \frac{\hat\omega^2}{(d-1)^2}\frac{u^{\frac{2d-8}{d-3}} (1-u^{\frac{d-1}{d-3}})}{(1-u)^2}\nonumber\\
& &
- (d-3)\frac{i\hat\omega}{d-1} \frac{[ d-4-(2d-5)u^{\frac{d-1}{d-3}}]u^{\frac{d-5}{d-3}} }{1-u} - (d-3)^2 \frac{i\hat\omega}{d-1}\frac{u^{\frac{2d-8}{d-3}} (1-u^{\frac{d-1}{d-3}})}{(1-u)^2}\nonumber\sioeea
One may
solve this equation perturbatively by separating
\siobe (\mathcal{H}_0 + \mathcal{H}_1) F = 0 \sioee
where
\siobea\label{sioeqH0} \mathcal{H}_0 F &\equiv& \mathcal{A} F'' + \mathcal{B}_0 F' + \mathcal{C}_{0 , 0} F \nonumber\\
\mathcal{H}_1 F &\equiv& (\mathcal{B}_{\hat\omega} - \mathcal{B}_0) F' + (\mathcal{C}_{\hat\omega , \hat L} - \mathcal{C}_{0 , 0}) F \nonumber\sioeea
Expanding the wavefunction perturbatively,
\siobe F = F_0 + F_1 + \dots \sioee
at zeroth order the wave equation reads
\siobe\label{sioeq22} \mathcal{H}_0 F_0 = 0 \sioee
whose acceptable solution is
\siobe\label{sioeq23} F_0 = u^{\frac{d-2}{2(d-3)}} \sioee
being regular at both the horizon ($u=1$) and the boundary ($u=0$, or $\Psi \sim r^{-\frac{d-2}{2}}\to 0$ as $r\to\infty$).
The Wronskian is
\siobe \mathcal{W} = \frac{1}{u^{\frac{d-4}{d-3}} (1-u^{\frac{d-1}{d-3}})} \sioee
and another linearly independent solution is
\siobe\label{sioeqchF0} \check F_0 = F_0\int \frac{\mathcal{W}}{F_0^2} \sioee
which is unacceptable because it diverges at both the horizon ($\check F_0 \sim \ln (1-u)$ for $u\approx 1$) and the boundary ($\check F_0 \sim u^{-\frac{d-4}{2(d-3)}}$ for $u\approx 0$, or $\Psi \sim r^{\frac{d-4}{2}} \to\infty$ as $r\to\infty$).

At first order the wave equation reads
\siobe \mathcal{H}_0 F_1 =- \mathcal{H}_1 F_0 \sioee
whose solution may be written as
\siobe\label{sioeqF1} F_1 = F_0\int \frac{\mathcal{W}}{F_0^2} \int \frac{F_0\mathcal{H}_1 F_0}{\mathcal{A}\mathcal{W}} \sioee
The limits of the inner integral may be adjusted at will
because this amounts to adding an arbitrary amount of the unacceptable solution.
To ensure regularity at the horizon, choose one of the limits of integration at $u=1$
rendering the integrand regular at the horizon.
Then at the boundary ($u=0$),
\siobe F_1 = \check F_0 \int_0^1 \frac{F_0\mathcal{H}_1 F_0}{\mathcal{A}\mathcal{W}} + \mathrm{regular~terms} \sioee
The coefficient of the singularity ought to vanish,
\siobe\label{sioeq29} \int_0^1 \frac{F_0 \mathcal{H}_1 F_0}{\mathcal{A}\mathcal{W}} = 0 \sioee
which yields a constraint on the parameters (dispersion relation)
\siobe\label{sioeqdisp} \mathbf{a}_0 \hat L^2 -i \mathbf{a}_1 \hat\omega - \mathbf{a}_2 \hat\omega^2 = 0 \sioee
After some algebra, one arrives at
\siobe\label{sioeqcoef} \mathbf{a}_0 = \frac{d-3}{d-1} \ \ , \ \ \ \
\mathbf{a}_1 = d-3 \sioee
The coefficient $\mathbf{a}_2$
may also be found explicitly for each dimension $d$,
but it cannot be written as a function of $d$ in closed form.
It does not contribute to the dispersion relation at lowest order.
E.g., for $d=4,5$, one obtains, respectively
\siobe\label{sioeqa2} \mathbf{a}_2 = \frac{65}{108} -\frac{1}{3}\ln 3 \ \ , \ \ \ \
\frac{5}{6}-\frac{1}{2}\ln 2 \sioee
Eq.~(\ref{sioeqdisp}) is quadratic in $\hat\omega$ and has two solutions,
\siobe \hat\omega_0 \approx -i\frac{\hat L^2}{d-1} \ \ , \ \ \ \  \hat\omega_1 \approx -i \frac{d-3}{\mathbf{a}_2} + i\frac{\hat L^2}{d-1} \sioee
In terms of the frequency $\omega$ and the quantum number $\ell$,
\siobe\label{sioeq34} \omega_0 \approx -i\frac{\ell(\ell+d-3)}{(d-1)r_+} \ \ , \ \ \ \  \frac{\omega_1}{r_+} \approx -i \frac{d-3}{\mathbf{a}_2} + i\frac{\ell(\ell+d-3)}{(d-1)r_+^2} \sioee
The smaller of the two, $\omega_0$,
is inversely proportional to the radius of the horizon
and is not included in the asymptotic spectrum.
The other solution, $\omega_1$,
is a crude estimate of the first overtone in the asymptotic spectrum, nevertheless
%
%
it shares two important features with the asymptotic spectrum:
it is proportional to $r_+$
and its dependence on $\ell$ is $\mathcal{O} (1/r_+^2)$.
The approximation may be improved by including higher-order terms.
This increases the degree of the polynomial in the dispersion relation (\ref{sioeqdisp}) whose roots then yield approximate values of more QNMs.
This method reproduces the asymptotic spectrum derived earlier albeit not in an efficient way.

To include finite size effects,
I shall use perturbation theory (assuming $1/r_+$ is small) and replace
$\mathcal{H}_1$ by
\siobe\label{sioeqH1} \mathcal{H}_1' = \mathcal{H}_1 + \frac{1}{r_+^2} \mathcal{H}_+ \sioee
where
\siobe \mathcal{H}_+ F \equiv \mathcal{A}_+ F'' + \mathcal{B}_+ F' + \mathcal{C}_+ F \sioee
The coefficients may be easily deduced by collecting $\mathcal{O} (1/r_+^2)$ terms in the exact wave equation.
One obtains
\siobea \mathcal{A}_+ &=& -2(d-3)^2 u^2(1-u) \nonumber\\
\mathcal{B}_+ &=& -(d-3) u\left[ (d-3)(2-3u) - (d-1) \frac{1-u}{1-u^{\frac{d-1}{d-3}}} u^{\frac{d-1}{d-3}} \right] \nonumber\\
\mathcal{C}_+ &=& \frac{d-2}{2} \left[ d-4-(2d-5)u - (d-1) \frac{1-u}{1-u^{\frac{d-1}{d-3}}} u^{\frac{d-1}{d-3}} \right] \nonumber\sioeea
Interestingly, the zeroth order wavefunction $F_0$ is an eigenfunction of $\mathcal{H}_+$,
\siobe \mathcal{H}_+ F_0 = -(d-2) F_0 \sioee
therefore the first-order finite-size effect is a simple shift of the angular momentum operator
\siobe \hat L^2 \to \hat L^2 - \frac{d-2}{r_+^2} \sioee
The QNMs of lowest frequency are modified to
\siobe\label{sioeqo0} \omega_0 = - i \frac{\ell(\ell+d-3)-(d-2)}{(d-1)r_+} + \mathcal{O} (1/r_+^2) \sioee
For $d=4, 5$, we have respectively,
\siobe \omega_0 = - i \frac{(\ell-1)(\ell+2)}{3r_+} \ \ , \ \ \ \  - i \frac{(\ell+1)^2-4}{4r_+} \sioee
in agreement with numerical results \cite{siob-CKL,siob-Princ}.

One deduces from (\ref{sioeqo0}) the maximum lifetime of the vector modes,
\siobe \tau_{\mathrm{max}} = \frac{4\pi}{d}\, T_H \sioee
In the case of a flat horizon ($\siopK = 0$),
\siobe \omega_0 = -i \frac{k^2}{(d-1)r_+} \sioee
which leads to the 
diffusion constant
\siobe D = \frac{1}{4\pi T_H} \sioee
In the case of a hyperbolic horizon ($\siopK = -1$), a similar calculation yields \cite{siob-AS1}
\siobe\label{sioVsoln}
\omega_0
= -i \frac{\xi^2 + \frac{(d-1)^2}{4}}{(d-1)r_+} \ \ ,
\ \ \tau = \frac{1}{|\omega_0|} < \frac{16\pi}{(d-1)^2}\, T_H \sioee
It follows that for $d=5$, these modes live longer than their spherical counterparts which is important for plasma behavior.

\subsection{Scalar perturbations}

Next I consider scalar perturbations which are calculationally more involved but phenomenologically more important because their spectrum contains the lowest frequencies and therefore the longest living modes.
For a scalar perturbation we ought to replace the potential $\hat V_{\mathsf{V}}$ by
\siobea  \hat V_{\mathsf{S}}(u) &=& \frac{\hat f(u)}{4} \left[ \hat m + \left( 1 + \frac{1}{r_+^2} \right) u\right]^{-2}  \nonumber\\
&\times& \Bigg\{ d(d-2) \left( 1+ \frac{1}{r_+^2} \right)^2 u^{\frac{2d-8}{d-3}}
- 6(d-2)(d-4)\hat m \left( 1+ \frac{1}{r_+^2} \right) u^{\frac{d-5}{d-3}}\nonumber\\
&& + (d-4)(d-6)\hat m^2u^{-\frac{2}{d-3}} +
(d-2)^2 \left( 1+ \frac{1}{r_+^2} \right)^3 u^3 \nonumber\\
&& + 2(2d^2-11d+18)\hat m \left( 1+ \frac{1}{r_+^2} \right)^2 u^2\nonumber\\
&& + \frac{(d-4)(d-6)\left( 1+\frac{1}{r_+^2} \right)^2}{r_+^2} u^2
- 3(d-2)(d-6)\hat m^2 \left( 1+\frac{1}{r_+^2} \right) u\nonumber\\
&& - \frac{6(d-2)(d-4)\hat m\left( 1+\frac{1}{r_+^2} \right)}{r_+^2} u + 2 (d-1)(d-2)\hat m^3 + d(d-2) \frac{\hat m^2}{r_+^2} \Bigg\} \nonumber\\ \sioeea
where
$\hat m = 2\frac{\ell (\ell+d-3) - (d-2)}{(d-1)(d-2)r_+^2} = \frac{2(\ell + d-2)(\ell -1)}{(d-1)(d-2)r_+^2}$.

In the large black hole limit $r_+\to \infty$ with $\hat m$ fixed (small), the potential simplifies to
\siobea  \hat V_{\mathsf{S}}^{(0)}(u) &=& \frac{1-u^{\frac{d-1}{d-3}}}{4( \hat m + u)^2}
\Bigg\{ d(d-2) u^{\frac{2d-8}{d-3}}
- 6(d-2)(d-4)\hat m u^{\frac{d-5}{d-3}}\nonumber\\
&& + (d-4)(d-6)\hat m^2u^{-\frac{2}{d-3}} +
(d-2)^2 u^3\nonumber\\
&&
+ 2(2d^2-11d+18)\hat m u^2
- 3(d-2)(d-6)\hat m^2  u
+ 2 (d-1)(d-2)\hat m^3  \Bigg\} \nonumber\\ \sioeea
The wave equation has an additional singularity due to the double pole of the scalar potential
at $u = -\hat m$.
It is desirable to factor out the behavior not only at the horizon, but also at the boundary and the pole of the scalar potential,
\siobe\label{sioeqPsiF} \Psi = (1-u)^{-i\frac{\hat\omega}{d-1}} \frac{u^{\frac{d-4}{2(d-3)}}}{\hat m + u} F(u) \sioee
Then the wave equation reads
\siobe\label{sioeqwsc} \mathcal{A} F'' + \mathcal{B}_{\hat\omega} F' + \mathcal{C}_{\hat\omega} F = 0 \sioee
where
\siobea \mathcal{A} &=& - (d-3)^2 u^{\frac{2d-8}{d-3}} (1-u^{\frac{d-1}{d-3}}) \nonumber\\
\mathcal{B}_{\hat\omega} &=& - (d-3) u^{\frac{2d-8}{d-3}} (1-u^{\frac{d-1}{d-3}}) \left[ \frac{d-4}{u} -\frac{2(d-3)}{\hat m + u} \right] \nonumber\\
&& - (d-3) [ d-4-(2d-5)u^{\frac{d-1}{d-3}}]u^{\frac{d-5}{d-3}} - 2(d-3)^2 \frac{i\hat\omega}{d-1}\frac{u^{\frac{2d-8}{d-3}} (1-u^{\frac{d-1}{d-3}})}{1-u} \nonumber\\
\mathcal{C}_{\hat\omega} &=&  - u^{\frac{2d-8}{d-3}} (1-u^{\frac{d-1}{d-3}}) \left[ -\frac{(d-2)(d-4)}{4 u^2} - \frac{(d-3)(d-4)}{u(\hat m + u)} + \frac{2(d-3)^2}{(\hat m + u)^2} \right] \nonumber\\
&& - \left[ \left\{ d-4-(2d-5)u^{\frac{d-1}{d-3}} \right\} u^{\frac{d-5}{d-3}} + 2(d-3) \frac{i\hat\omega}{d-1}\frac{u^{\frac{2d-8}{d-3}} (1-u^{\frac{d-1}{d-3}})}{1-u}\right]\left[ \frac{d-4}{2u} - \frac{d-3}{\hat m + u} \right] \nonumber\\
&& - (d-3)\frac{i\hat\omega}{d-1} \frac{[ d-4-(2d-5)u^{\frac{d-1}{d-3}}]u^{\frac{d-5}{d-3}} }{1-u} - (d-3)^2 \frac{i\hat\omega}{d-1}\frac{u^{\frac{2d-8}{d-3}} (1-u^{\frac{d-1}{d-3}})}{(1-u)^2}\nonumber\\
&& + \frac{\hat V_{\mathsf{S}}^{(0)}(u)-\hat\omega^2}{1-u^{\frac{d-1}{d-3}}} + (d-3)^2 \frac{\hat\omega^2}{(d-1)^2}\frac{u^{\frac{2d-8}{d-3}} (1-u^{\frac{d-1}{d-3}})}{(1-u)^2}\nonumber\sioeea
I shall define the zeroth-order wave equation as $\mathcal{H}_0 F_0 = 0$, where
\siobe\label{sioeq0sc} \mathcal{H}_0 F \equiv \mathcal{A} F'' + \mathcal{B}_0 F' \sioee
The acceptable zeroth-order solution is
\siobe\label{sioeqF0sc} F_0(u) = 1 \sioee
which is plainly regular at all singular points ($u=0,1, -\hat m$).
It corresponds to a wavefunction vanishing at the boundary
($\Psi \sim r^{-\frac{d-4}{2}}$ as $r\to\infty$).

The Wronskian is
\siobe\label{sioeqWsc} \mathcal{W} = \frac{\left( \hat m + u\right)^2 }{u^{\frac{2d-8}{d-3}} (1-u^{\frac{d-1}{d-3}})} \sioee
and an nacceptable solution is $ \check F_0 = \int \mathcal{W} $.
It can be written in terms of hypergeometric functions.
For $d\ge 6$, it has a singularity at the boundary, $\check F_0 \sim u^{-\frac{d-5}{d-3}}$ for $u\approx 0$,
or $\Psi \sim r^{\frac{d-6}{2}}\to\infty$ as $r\to\infty$.
For $d=5$, the acceptable wavefunction behaves as $r^{-1/2}$ whereas the unacceptable one behaves as $r^{-1/2} \ln r$.
For $d=4$, the roles of $F_0$ and $\check F_0$ are reversed, however the results still valid because the correct boundary condition at the boundary is a Robin boundary condition \cite{siob-ego,siob-MP}.
Finally, note that $\check F_0$ is also singular (logarithmically) at the horizon ($u=1$).

Working as in the case of vector modes,
one arrives at the first-order constraint
\siobe\label{sioeqcnssc} \int_0^1 \frac{\mathcal{C}_{\hat\omega}}{\mathcal{A}\mathcal{W}} = 0 \sioee
because $\mathcal{H}_1 F_0 \equiv (\mathcal{B}_{\hat\omega} - \mathcal{B}_0) F_0'
+ \mathcal{C}_{\hat\omega} F_0 = \mathcal{C}_{\hat\omega} $.
This leads to the
dispersion relation
\siobe\label{sioeqcnssc2} \mathbf{a}_0 - \mathbf{a}_1 i\hat\omega - \mathbf{a}_2 \hat\omega^2 = 0 \sioee
After some algebra, one obtains
\siobe \mathbf{a}_0 = \frac{d-1}{2} \ \frac{ 1+ (d-2)\hat m}{(1+ \hat m )^2} \ \ ,
\ \ \ \ \mathbf{a}_1 = \frac{d-3}{(1+ \hat m )^2} \ \ ,
\ \ \ \ \mathbf{a}_2 = \frac{1}{\hat m} \left\{ 1 + O(\hat m) \right\} \sioee
For small $\hat m$, the quadratic equation has solutions
\siobe\label{sioeqosc1} \hat\omega_0^\pm \approx - i\frac{d-3}{2} \ \hat m \pm \sqrt{ \frac{d-1}{2} \ \hat m} \sioee
related to each other by $\hat\omega_0^+ = -\hat\omega_0^{-*}$, which is a
general symmetry of the spectrum.


Finite size effects
at first order amount to a shift of the coefficient $\mathbf{a}_0$ in the dispersion relation
\siobe \mathbf{a}_0 \to \mathbf{a}_0 + \frac{1}{r_+^2} \mathbf{a}_+ \sioee
After some tedious but straightforward algebra, we obtain
\siobe \mathbf{a}_+ = \frac{1}{\hat m} \left\{ 1 + O(\hat m) \right\} \sioee
The modified dispersion relation yields the modes
\siobe\label{sioeqosc2} \hat\omega_0^\pm \approx - i\frac{d-3}{2} \ \hat m \pm \sqrt{ \frac{d-1}{2} \ \hat m +1} \sioee
In terms of the quantum number $\ell$,
\siobe\label{sioeqosc2a} \omega_0^\pm \approx - i(d-3) \ \frac{\ell (\ell+d-3)-(d-2)}{(d-1)(d-2) r_+} \pm \sqrt{ \frac{\ell (\ell+d-3)}{d-2}} \sioee
in agreement with numerical results \cite{siob-Princ}.


Notice that the imaginary part is inversely proportional to $r_+$, as in vector case.
In the scalar case, we also obtained a
finite real part independent of $r_+$.

The maximum lifetime of a gravitational scalar mode is found from (\ref{sioeqosc2a}) to be
\siobe \tau_{\mathrm{max}} = \frac{d-2}{(d-3)d}\, 4\pi T_H \sioee
In the case of a flat horizon ($\siopK = 0$), one obtains
\siobe \omega=\pm\frac{k}{\sqrt{d-2}}-i\frac{d-3}{(d-1)(d-2)r_+}\ k^2 \sioee
showing that the speed of sound is
\siobe v = \frac{1}{\sqrt{d-2}}\sioee
as expected for a CFT and the diffusion constant is
\siobe D = \frac{d-3}{d-2}\ \frac{1}{4\pi T_H} \sioee
For a hyperbolic horizon ($\siopK = -1$), a similar calculation yields \cite{siob-AS1}
\siobe\label{sioVsoln2}
\omega
= \pm \sqrt{\frac{\xi^2 + (\frac{d-3}{2})^2}{d-2}} -i \frac{(d-3)[\xi^2 + \frac{(d-1)^2}{4}]}{(d-1)(d-2)r_+} \ \ ,
\ \ \tau < \frac{4(d-2)}{(d-3)(d-1)^2}\, 4\pi T_H \sioee
In the physically relevant case $d=5$, evidently the $\siopK=-1$ scalar modes live longer than any other modes, which is important for plasma behavior.

\subsection{Tensor perturbations}

Finally, for completeness I discuss the case of tensor perturbations.
Unlike the other two cases of gravitational perturbations, the asymptotic spectrum of tensor perturbations is the entire spectrum.
To see this,
note that in the large black hole limit, the wave equation reads
\siobea\label{sioeqwavt} - (d-3)^2 (u^{\frac{2d-8}{d-3}} -u^3)\Psi'' - (d-3) [ (d-4)u^{\frac{d-5}{d-3}}-(2d-5)u^2]\Psi' && \nonumber\\
 + \left\{ \hat L^2 + \frac{d(d-2)}{4}u^{-\frac{2}{d-3}} + \frac{(d-2)^2}{4} u - \frac{\hat\omega^2}{1-u^{\frac{d-1}{d-3}}} \right\} \Psi &=& 0 
\nonumber\sioeea
For the zeroth-order equation, we may set $\hat L = 0 = \hat\omega$.
The resulting equation may be solved exactly.
Two linearly independent solutions are ($\Psi = F_0$ at zeroth order)
\siobe F_0(u) = u^{\frac{d-2}{2(d-3)}} \ \ , \ \ \ \ \check F_0(u) = u^{-\frac{d-2}{2(d-3)}} \ln \left( 1-u^{\frac{d-1}{d-3}} \right) \sioee
Neither behaves nicely at both ends ($u=0,1$).
Therefore both are unacceptable which makes it
impossible to build a perturbation theory to calculate small frequencies which are inversely proportional to $r_0$.
This negative result is
in agreement with numerical results \cite{siob-CKL,siob-Princ} and in accordance with the
AdS/CFT correspondence.
Indeed,
there is no ansatz that can be built from tensor spherical harmonics $\mathbb{T}_{ij}$ satisfying the linearized hydrodynamic equations, because of the conservation and tracelessness properties of $\mathbb{T}_{ij}$.

\subsection{Hydrodynamics on the AdS boundary}

The above results in the bulk dictate the hydrodynamic behavior of the dual gauge theory fluid on the conformal boundary.
To see the correspondence, one needs to understand the hydrodynamics in the linearized regime of a $d-1$ dimensional fluid with dissipative effects.
The fluid lives on a space with
metric
\siobe ds_{\partial}^2=-dt^2+d\Sigma_{\siopK,d-2}^2\sioee
The hydrodynamic equations are simply the requirement that the stress-energy momentum tensor be conserved,
\siobe \nabla_\mu T^{\mu\nu}=0
\sioee
As the duality corresponds to a conformal field theory one must also demand scale invariance which implies
\siobe
T^{\mu}_{~\mu}=0 \ , \ \ \epsilon=(d-2)p \ \ , \ \ \zeta =0
\sioee
where $\epsilon$, $p$ and $\zeta$ are the energy density, pressure and bulk viscosity of the fluid.
%
In the rest frame of the fluid, the velocity field is $u^\mu=(1,0,0,0)$ and the pressure $p_0$ is constant. Consider a perturbation
\siobe
u^\mu=(1,u^i) \ \ , \ \ \ \ p=p_0+\delta p\sioee
Applying the hydrodynamic equations, one obtains
\siobes
(d-2)\partial_t \delta p+(d-1)p_0 \nabla_i u^i&=&0\nonumber\\
(d-1)p_0 \partial_t u^i +\partial^i\delta p-\eta\left[\nabla^j \nabla_j u^i+\siopK(d-3) u^i+\frac{d-4}{d-2}\partial^i (\nabla_j u^j)\right]&=&0
\sioees
where I used the curvature tensor $R_{ij}=\siopK(d-3)g_{ij}$.

For {\em vector perturbations}, consider the {\em ansatz}
\siobe
\delta p=0~,~~~u^i=\mathcal{C}_V e^{-i\omega t}{\mathbb V}^i
\sioee
where ${\mathbb V}^i$ is a vector harmonic.

The hydrodynamic equations imply
\siobe\label{sioVhydro}
-i \omega (d-1) p_0+\eta\left[k_V^2-\siopK(d-3)\right]=0
\sioee
Using
\siobe
\frac{\eta}{p_0}=(d-2)\frac{\eta}{s}\frac{S}{M}=\frac{4\pi\eta}{s}\frac{r_+}{\siopK+r_+^2}
\sioee
with $\omega$ from the gravity dual, one obtains
for large $r_+$,
\siobe
\frac{\eta}{s} =\frac{1}{4\pi}
\sioee
which is the standard value of the ratio in gauge theory fluids with a gravity dual \cite{siob-PSS}.

For {\em scalar perturbations}, consider the {\em ansatz}
\siobe
u^i=\mathcal{A}_S e^{-i\omega t}\partial^i \mathbb{S}~,~~~\delta p=\mathcal{B}_S e^{-i\omega t}\mathbb{S}
\sioee
where $\mathbb{S}$ is a scalar harmonic.

The hydrodynamic equations imply the system of equations
\siobes
&&(d-2)i\omega \mathcal{B}_S+(d-1)p_0 k_S^2\mathcal{A}_S=0\nonumber\\
&&\mathcal{B}_S+\mathcal{A}_S\left[-i\omega(d-1)p_0-2(d-3)\siopK\eta +2 \eta k_S^2\frac{d-3}{d-2} \right]=0
\sioees
The determinant must vanish,
\siobe \left| \begin{array}{cc} (d-2)i\omega & (d-1) p_0 k_S^2 \\
1 & -i\omega (d-1) p_0 - 2(d-3)\siopK\eta + 2\eta k_S^2 \,\frac{d-3}{d-2} \end{array} \right| = 0 \sioee
Arguing along the same lines as for vector perturbations, we arrive at
\siobe\label{sioShydro}
\frac{\eta}{s} =\frac{1}{4\pi}\sioee
which is the same result as the one obtained with vector QNMs.


\subsection{Conformal soliton flow}

The above results have been applied to the study of the quark-gluon plasma which forms in heavy ion collisions (at the Relativistic Heavy Ion Collider (RHIC) and elsewhere).
In the case of a spherical horizon
($\siopK =+1$),
%
the boundary of spacetime is $S^3\times\mathbb{R}$.
This may be conformally mapped onto a flat Minkowski space.
Then by holographic renormalization,
the AdS$_5$-Schwarzschild black hole is dual to a spherical shell of plasma on the four-dimensional Minkowski space which first contracts and then expands (conformal soliton flow) \cite{siob-Princ}.

Quasi-normal modes govern the properties of this plasma with long-lived modes (i.e., of small $\Im\omega$) having the most influence.
For example, one obtains the ratio
\siobe \frac{v_2}{\delta} = \frac{1}{6\pi} \Re \frac{\omega^4-40\omega^2+72}{\omega^3-4\omega} \sin \frac{\pi\omega}{2} \sioee
where
$v_2 = \langle \cos 2\phi \rangle$ evaluated at $\theta = \frac{\pi}{2}$ (mid-rapidity) and averaged with respect to the energy density at late times;
$\delta = \frac{\langle y^2 - x^2 \rangle}{\langle y^2 + x^2 \rangle}$ is the eccentricity at time $t=0$.
Numerically, $\frac{v_2}{\delta} = 0.37$, which compares well with the result from RHIC data, $\frac{v_2}{\delta} \approx 0.323$ \cite{siob-PHENIX}.

Another observable is the thermalization time which is found to be
\siobe \tau = \frac{1}{2|\Im \omega|} \approx \frac{1}{8.6 T_{\mathrm{peak}}} \approx 0.08~\mathrm{fm/c}\ \ , \ \ \ \ T_{\mathrm{peak}} = 300~\mathrm{MeV} \sioee
not in agreement with the RHIC result $\tau \sim 0.6$~fm/c \cite{siob-RHIC}, but still encouragingly small.
For comparison, the corresponding result from perturbative QCD is $\tau \gtrsim 2.5$~fm/c \cite{siob-QCD,siob-QCD2}.

In the case of a hyperbolic horizon (topological black hole;
$\siopK =-1$),
one needs to work with a conformal map from $\mathbb{H}^{d-2}/\Gamma \times \mathbb{R}$ to a $(d-1)-$dimensional Minkowski space.
Finding an explicit form of this map for $d=5$ involves a considerable amount of numerical work. However,
it is important that one consider this case because the modes of hyperbolic black holes live the longest \cite{siob-AS1}.



\section{Phase transitions}
\label{siosec:4}





In this section I discuss hairy black holes in asymptotically AdS space and their duals. At low temperatures, an instability leads to symmetry breaking and the formation of a dual superconductor. ELectromagnetic perturbations of the black hole determine the conductivity in the bulk. First I review the case of a flat horizon ($\siopK=0$) \cite{siob-HHH} and then I discuss the case of hyperbolic horizon ($\siopK=-1$) where exact analytical results are obtained \cite{siob-SYN}.

\subsection{$\siopK=0$}

Consider a scalar $\Psi$ of mass $m^2 = -2$, which is above the Breitenlohner-Freedman (BF) bound coupled to an electromagnetic potential $A_\mu$ in 3+1 dimensions. The Lagrangian density is
\siobe\label{sioLN1} \mathcal{L} = -\frac{1}{2} \partial_\mu \Psi \partial^\mu \Psi + \Psi^2 - \frac{1}{4} F_{\mu\nu} F^{\mu\nu} - \frac{q^2}{2} \Psi^2 (\partial_\mu\theta - A_\mu) (\partial^\mu\theta - A^\mu) \sioee
where $\theta$ is a St\"uckelberg field. $q$ is an arbitrary parameter which can be throught of as the electric charge of the scalar field $\Psi$ (one may instead turn $\Psi$ into a complex scalar field of charge $q$ coupled to an electromagnetic potential in a standard fashion).

The Lagrangian density (\ref{sioLN1}) is invariant under the $U(1)$ gauge transformation
\siobe A_\mu \to A_\mu + \partial_\mu\omega \ , \ \ \theta\to\theta + \omega \sioee
To fix the gauge, set
\siobe \theta = 0 \sioee
Working in the probe limit ($q\to\infty$) in which there is no back reaction to the metric, assume that the fields propagate
in the black hole
background (\ref{siometric}) with $d=4$ and $\siopK=0$.
The radius of the horizon and Hawking temperature are, respectively,
\siobe r_+ = (2\mu)^{1/3} \ , \ \ T = \frac{3r_+}{4\pi} \sioee
Assuming spherical symmetry and an electrostatic potential $A_0 = \Phi(r)$, the field equations yield two coupled non-linear differential equations \cite{siob-HHH}
\siobes \Psi'' + \left( \frac{f'}{f} + \frac{2}{r} \right) \Psi' + \left( \frac{\Phi}{f} \right)^2 \Psi + \frac{2}{f} \Psi &=& 0 \nonumber\\
\Phi'' + \frac{2}{r} \Phi' - \frac{2\Psi^2}{f} \Phi &=& 0 \sioees
where I set $q=1$ and
\siobe f(r) = r^2 - \frac{2\mu}{r} \sioee
As $r\to\infty$, one obtains the boundary behavior
\siobe \Psi = \frac{\Psi^{(1)}}{r} + \frac{\Psi^{(2)}}{r^2} + \dots \ , \ \
\Phi = \Phi^{(0)} + \frac{\Phi^{(1)}}{r} + \dots \sioee
where one of the $\Psi^{(i)}=0$ ($i=1,2$) for stability, $\Phi^{(0)}$ is the chemical potential and $\Phi^{(1)} = -\rho$ (charge density).

Below a critical temperature $T_0$ a condensate forms,
\siobe \langle \mathcal{O}_i \rangle = \sqrt{2} \Psi^{(i)} \sioee
of an operator of dimension $\Delta = i$.

At $T=T_0$, one may set $\Psi=0$ in the equation for $\Phi$ and deduce (using $\Phi (r_+) = 0$)
\siobe \Phi = \rho \left( \frac{1}{r_+} - \frac{1}{r} \right) \sioee
Then the equation for $\Psi$ turns into an eigenvalue problem which yields
\[ T_0 \approx 0.226 \sqrt\rho \ , \ \ 0.118 \sqrt\rho \]
depending on the boundary conditions.

To study the properties of the dual CFT, apply an electromagnetic perturbation.
It obeys the wave equation
\siobe\label{sioEM1} A'' + \frac{f'}{f} A' + \left( \frac{\omega^2}{f^2} - \frac{2\Psi^2}{f} \right) A = 0 \sioee
to be solved subject to the boundary conditions that it be
ingoing at the horizon, $A\sim f^{-i\omega/(4\pi T)}$, and at the boundary ($r\to\infty$),
\siobe A = A^{(0)} + \frac{A^{(1)}}{r} + \dots \sioee
Ohm's law yields the conductivity
\siobe \sigma (\omega) = \frac{A^{(1)}}{i\omega A^{(0)}} \sioee
For $T\ge T_0$, $\Psi = 0$, therefore $A\sim e^{i\omega r_*}$ where $r_* = \int dr/f(r)$ is the tortoise coordinate. It follows that
\siobe \sigma (\omega) = 1 \sioee
%
At low $T$, for $\langle\mathcal{O}_1\rangle \ne 0$, we have
\[ \Psi \approx  \frac{\langle\mathcal{O}_1 \rangle}{\sqrt{2}\ r} \]
Since $r_+\to 0$, we obtain $A\sim e^{i\omega' r_*}$, where $\omega' = \sqrt{\omega^2 - \langle\mathcal{O}_1\rangle^2}$.
Therefore, for $\omega < \langle\mathcal{O}_1\rangle$, $\Re\sigma = 0$, i.e., we obtain a
superconductor with a gap.


\subsection{$\siopK=-1$}

Turning to the case of a hyperbolic horizon \cite{siob-SYN}, choose a
%
scalar $\Psi$ of mass $m^2 = -2$, as before, but conformally coupled with potential
\[ V(\Psi) = \frac{8\pi G}{3} \Psi^4 \]
The system has an exact solution
 (MTZ black hole \cite{siob-MTZ})
\siobe
d{s}^{2}=-f_{MTZ} (r) dt^{2}+
\frac{dr^{2}}{f_{MTZ} (r)} +r^{2}d\sigma^{2} \quad, \quad f_{MTZ} = r^{2}-\biggl(1+\frac{r_0}{r}\biggr)^{2}~,
\sioee
with
\siobe\label{sioPsiconf1} \Psi (r) = -\sqrt{\frac{3}{4\pi
G}}\frac{r_0}{r+r_0} \ \ , \ \ \ \ \Phi = 0 \sioee
The
temperature, entropy and mass
are, respectively, \siobe T=\frac{1}{
\pi} \left( r_+ - \frac{1}{2} \right)~,\quad S_{MTZ}=\frac{\sigma }{4G} \left( 2r_{+}-1 \right)~,\quad
M_{MTZ}=\frac{\sigma r_{+}}{4 \pi
G} \left( r_+ - 1 \right)~.\label{siorelations1} \sioee
and the law of
thermodynamics $dM=TdS$ holds.

At $M=0$, the MTZ black hole coincides with the topological black hole with no hair (eq.~(\ref{siometric}) with $d=4$, $\siopK=-1$),
\siobe
ds_{\mathrm{AdS}}^{2}=-( r^{2} -1) dt^{2}+\frac{dr^2}{r^{2}-1}
+r^{2}d\Sigma^{2}
\sioee
and an enhanced scaling symmetry (pure AdS space) emerges at the critical temperature
\siobe\label{sioeqTcr} T_0 = \frac{1}{2\pi } \sioee
At this point there is a phase transition which can be seen by calculating the difference in free energies,
\siobe \Delta F=
F_{TBH}-F_{MTZ}=-\frac{\sigma}{8\pi G}\ \pi^3 l^3
(T-T_{0})^{3}+\dots~,\label{siodifference} \sioee
showing that there is a third-order
phase transition at
$T_0$.

Perturbative stability of the MTZ black hole has also been demonstrated for $T<T_0$ ($M<0$) \cite{siob-SYN}. Comparing with the flat case, note that here both $\Psi^{(1)}$ and $\Psi^{(2)}$ are non-vanishing, yet the MTZ black hole is stable. However this is true only if the mass is negative which is never the case with a flat horizon.
%
%
Also, here the condensation of the scalar field has a geometrical
origin and is due entirely to its coupling to gravity.

Moreover, the heat capacities in the normal and superconducting (corresponding to the MTZ black hole) phases, respectively,
as $T\to 0$ exhibit a power-law behavior
\siobe\label{sioeqheatc} C_n \approx
\frac{\pi\sigma }{3\sqrt 3 G}\ T \ \ , \ \ \ \ C_s \approx
\frac{\pi\sigma }{2G}\ T \ , \sioee
Since both $\Psi^{(1)}$ and $\Psi^{(2)}$ are non-vanishing, we have a multi-trace deformation of the CFT \cite{siob-MT} with a condensate
\siobe \langle
\mathcal{O}_{1}\rangle = \sqrt{\frac{3\pi^3}{2 G}}\, (T_0^2 -T^2)
\,. \sioee
It should be noted that the deformation does not break the global $U(1)$ symmetry because $\Psi$ is a real field (see eq.~(\ref{sioLN1})).


To study the conductivity, apply an electromagnetic perturbation.
It obeys the wave equation (\ref{sioEM1}) which may be solved using first-order
%
perturbation theory in $q^2$,
\siobe A=e^{- i\omega
r_{*}}+\frac{q^{2}}{2i\omega}e^{ i\omega
r_{*}}\int^{r}_{r_{+}}dr'\Psi^{2}(r')e^{-2 i\omega r_{*}}
-\frac{q^{2}}{2i\omega}e^{- i\omega
r_{*}}\int^{r}_{r_{+}}dr'\Psi^{2}(r')~.\sioee
The conductivity to
first order in $q^{2}$ is
\siobe
\sigma(\omega)=\frac{A^{(1)}}{i\omega
A^{(0)}}=1-\frac{q^{2}}{i\omega}\int^{\infty}_{r_{+}}dr\Psi^{2}(r)
e^{-2 i\omega r_{*}}~.\label{sioconductivitymtz}\sioee
The superfluid
density is found from
\siobe \Re[\sigma(\omega)] \sim \pi n_s
\delta(\omega) \ \ , \ \ \ \
\Im[\sigma(\omega)] \sim \frac{n_s}{\omega} ~,\quad
\omega \to 0~. \sioee
One obtains
\siobe n_{s}=q^{2}\int^{\infty}_{r_{+}}dr\Psi^{2}(r)=\frac{3q^{2}}{4\pi
G}\, \frac{r_0^{2} }{r_{+}+r_0 } = \alpha\left(T_{0}-T\right)^{2} ~,\quad \alpha = \frac{3\pi
q^{2}}{4G}~. \sioee
Near $T=0$,
\siobe\label{sioeqnsa} n_s(0) - n_s(T) \approx \frac{\alpha}{\pi} T^\delta \ , \ \ \delta =1 \sioee
In table \ref{siotable2}, this analytic prediction is compared against exact numerical results for various values of the charge $q$. Naturally, the agreement is best at small values of $q$.
\begin{table}
\begin{center}

\begin{tabular}{||c||c|c|c||}

\hline

$q/\sqrt G$ & $1$ & $3$&
$5$ \\

\hline

$\delta$ & $1.025\pm 0.007$ &  $1.52\pm 0.03$ & $1.78\pm 0.03$ \\

\hline

\end{tabular}

\end{center}
\caption{The exponent $\delta$ characterizing the low-temperature dependence of the superfluid density $n_s$.
}\label{siotable2}
\end{table}

The
normal, non-superconducting,
component of the DC conductivity is
\siobe n_n = \lim_{\omega \to 0} \Re
[\sigma(\omega)] \,. \label{sionormalConductivity}\sioee
Therefore,
\siobe
\ln n_{n}=2q^{2}\int^{\infty}_{r_{+}}dr\Psi^{2}(r)r_{*}~. \sioee
At low $T$,
\siobe n_{n}\sim
T^{\gamma}\,, \,\,\,\,\, \gamma=\frac{3q^{2}}{4\pi
G}~.\label{siopolygap} \sioee
This analytic result and the prediction for the parameter $\alpha$ determining the critical behavior of the superfluid density are compared against exact numerical results in table \ref{siotable1}.
Again, the agreement is best at small $q$.
\begin{table}
\begin{center}

\begin{tabular}{||c|c|c|c|c||}

\hline

$q/\sqrt G$ & $\gamma_{\mathrm{numerical}}$ & $\gamma_{\mathrm{analytical}}$&
$\alpha_{\mathrm{numerical}}$ & $\alpha_{\mathrm{analytical}}$ \\

\hline

0.1 & 0.0020 &  0.0024 & 0.0225 & 0.024 \\

0.5 & 0.0538 &  0.0597 & 0.552 & 0.589 \\

1.0 & 0.187 &  0.239 & 2.196 & 2.356 \\

2.0 & 0.684 &  0.955 & 8.678 & 9.425 \\

3.0 & 1.325 &  2.15 & 20.35 & 21.21 \\

5.0 & 2.522 &  5.97 & 52.90 & 58.90 \\

\hline

\end{tabular}
\end{center}
\caption{Numerical {\it vs}~analytical results for the normal and superfluid densities.
}\label{siotable1}
\end{table}

\begin{figure}[!t]
\begin{center}
\includegraphics[scale=0.22,angle=-90]{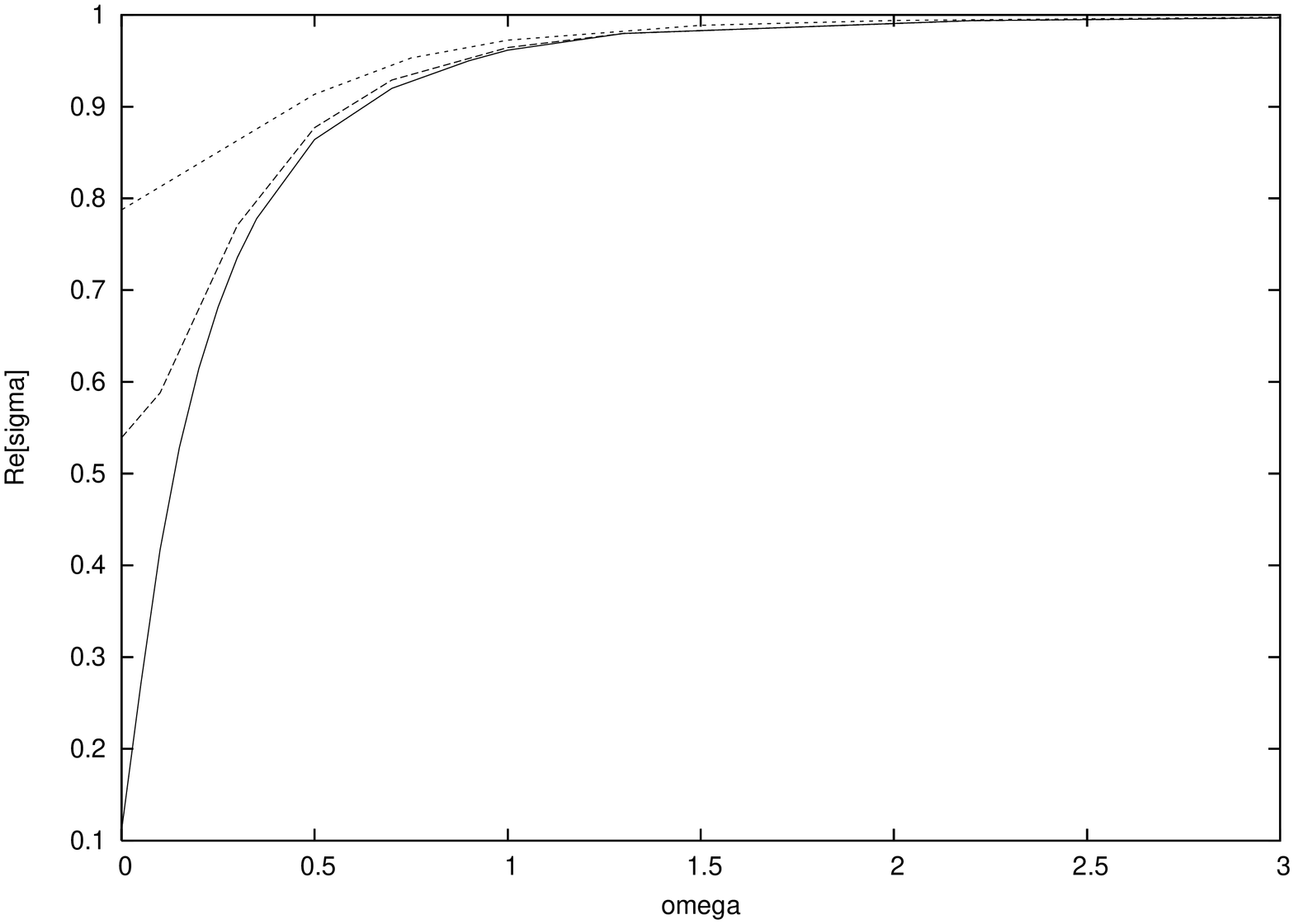}
\includegraphics[scale=0.22,angle=-90]{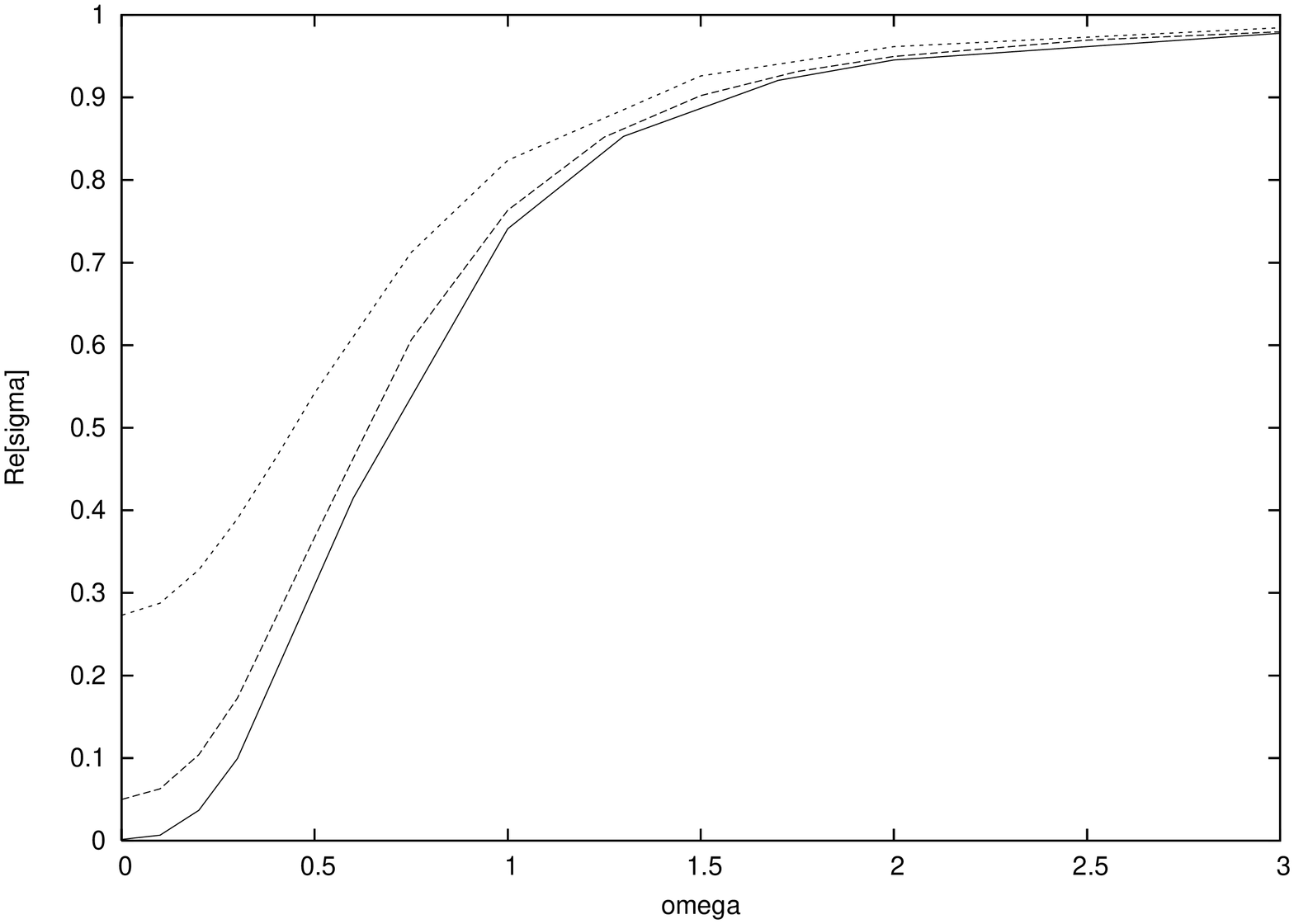}
\end{center}

\caption{The real part of the conductivity {\em vs} $\omega$
for $q/\sqrt G=2$ (left) and $q/\sqrt G=5$ (right) and $T=0.0032, 0.032, 0.064.$ The lowest curve
corresponds to the lowest temperature.} \label{siore_q}
\end{figure}

\begin{figure}[!t]
\begin{center}
\includegraphics[scale=0.22,angle=-90]{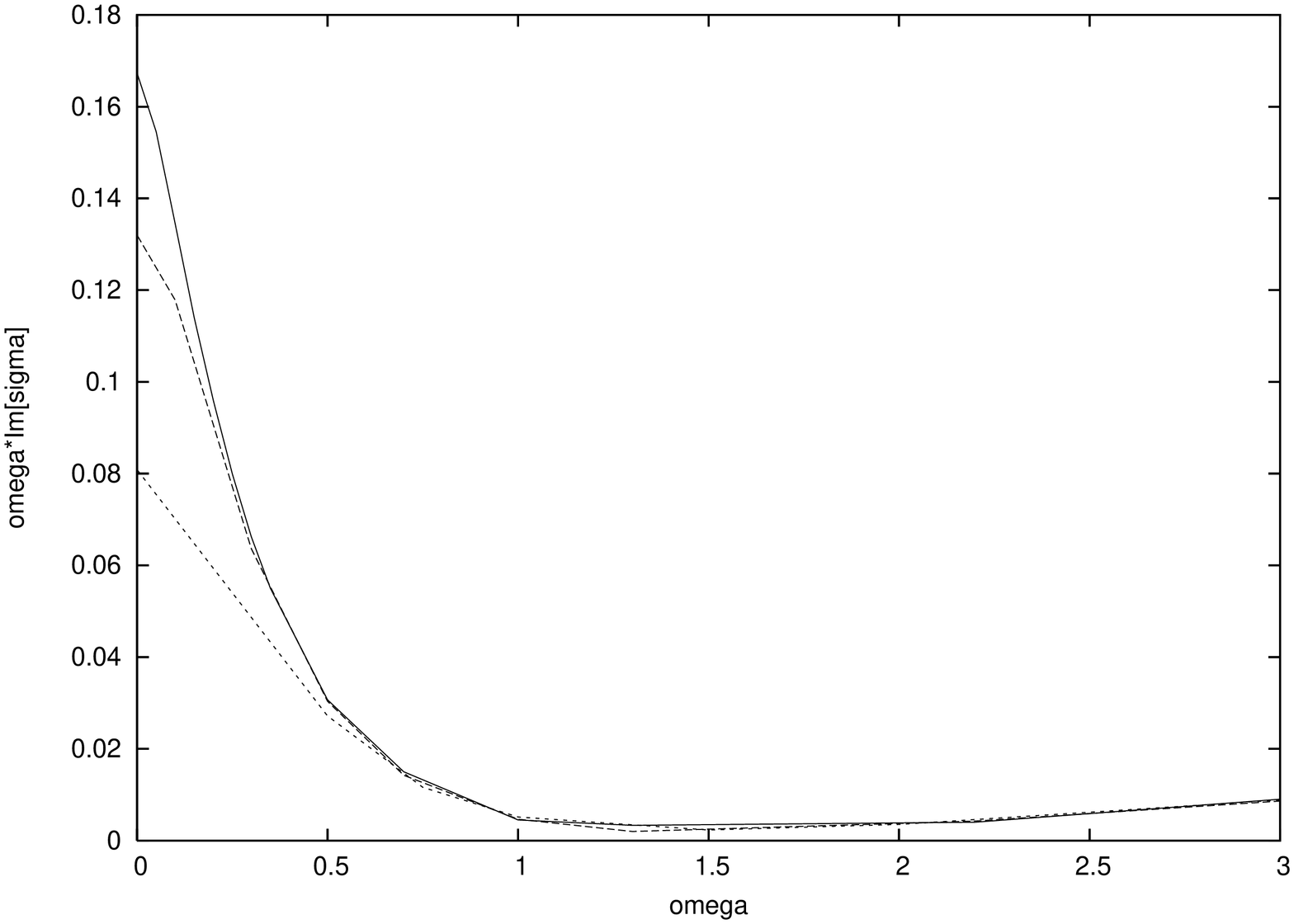}
\includegraphics[scale=0.22,angle=-90]{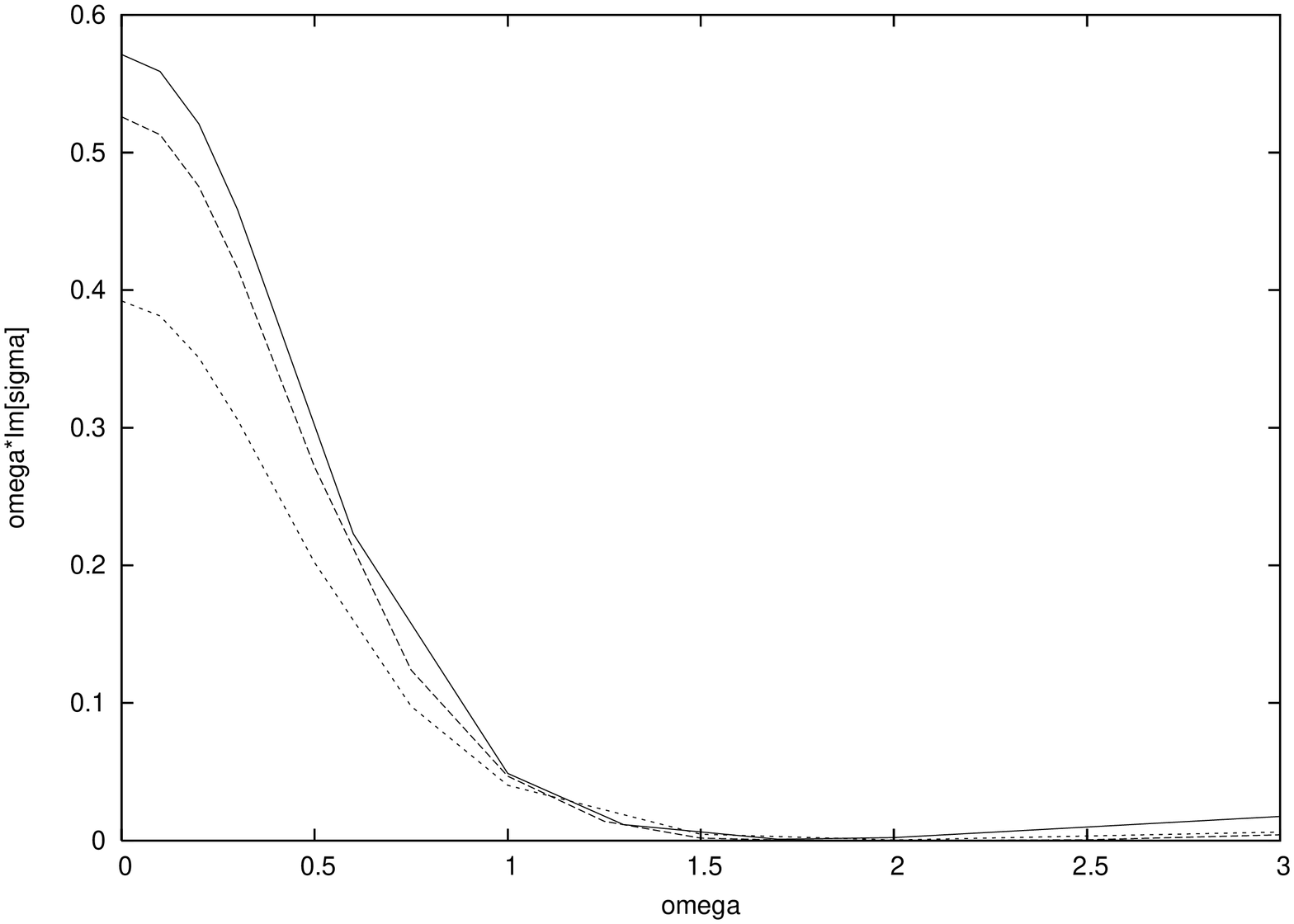}
\end{center}

\caption{The imaginary part of the conductivity multiplied by
$\omega$ {\em vs} $\omega$ for $q/\sqrt G=2$ (left) and $q/\sqrt G=5$ (right) and $T=0.0032, 0.032,
0.064.$ The uppermost curve corresponds to the lowest
temperature.} \label{sioim_q}
\end{figure}
Figures \ref{siore_q} and \ref{sioim_q} show the frequency dependence of the real and imaginary, respectively, parts of the conductivity.
The real part of the conductivity becomes
smaller as we increase the charge $q$.
Unfortunately, numerical instabilities also increase and we have not been able to produce reliable numerical results above $q/\sqrt G =5$.
The superconductor appears to be {\em gapless}.
However, a gap is likely to develop above a certain value of the charge $q$, as
indicated by the trend in the graphs as $q$ increases.


\section{Conclusion}
\label{siosec:5}

The quasi-normal modes that govern perturbations of black holes in asymptotically AdS space are a powerful tool in understanding the hydrodynamic behavior of a gauge theory fluid at strong coupling.
Here I focused on the analytic calculation of QNMs. I discussed both high overtones and low frequencies. I applied the results on gravitational perturbations to the understanding of the quark-gluon plasma produced in heavy ion collisions at RHIC and the LHC.
I also considered hairy black holes whose electromagnetic perturbations allow one to analyze the conductivity of the dual conformal field theory and the phase transition to a superconducting state.
I reviewed the case of a flat horizon and compared the results with those from black holes with hyperbolic horizon for which exact hairy solutions have been constructed (MTZ black holes \cite{siob-MTZ}).
In all these cases, only the low-lying QNMs were needed. It is unclear what physical role high overtones play.


%
%
%

\end{document}